\documentclass[slac_one]{revtex4}
\usepackage{graphicx}
\usepackage{fancyhdr}
\begin{document}

\title{The Liquid Argon Jet Trigger of the H1 Experiment at HERA} 

%

\author{Bob Olivier}
\affiliation{Max-Planck-Institut f\"ur Physik (Werner-Heisenberg-Institut), F\"ohringer Ring 6, D-80805 M\"unchen, Germany}

\begin{abstract}
The Liquid Argon Jet Trigger, installed in the
H1 experiment\index{H1 experiment} at HERA\index{HERA},
implements in $800$~ns a real-time cluster algorithm by finding local
energy maxima, summing their immediate neighbors, sorting the
resulting ``jets'' by energy, and applying topological conditions. It
operated since the year $2006$ and drastically reduced the thresholds
for triggering on electrons and jets.
\end{abstract}

\maketitle

\thispagestyle{fancy}


\section{Introduction}
After the luminosity upgrade of the HERA machine in the years
$2000-2001$ (HERA$-2$), a significant increase of the background rates
was expected and indeed observed. While parts of the H1 detector were
upgraded during the year $2001$ as well, the H1 data logging rate to
permanent storage (about $10$~Hz) remained a stringent constraint for
the data acquisition system. The aim of the upgrade of the digital
part of the LAr trigger, the Jet Trigger\index{Jet Trigger}~\cite{Dubak:2007zza}, 
was to complement the
existing global LAr calorimeter\index{calorimeter} trigger\index{trigger} 
with a system that performs
real-time clustering to avoid summing-up noise distributed over large
parts of the calorimeter, thus allowing for triggers on even lower
energy depositions while keeping the trigger rates within the
required bounds.

\section{Jet Trigger Algorithm}
The Jet Trigger identifies the localized energy depositions of
electrons, photons and bundles of hadrons in the LAr calorimeter, and
uses these energy clusters (``jets''), including their topological
information, for a fast event selection. The ``jets'' are found by
identifying trigger towers with a local energy maximum. Around this
maximum the immediate neighboring towers are summed and added to the
center. The resulting local ``jets'' are the basis of the trigger
decision. Such a local concept improves the sensitivity for
low-energy depositions in the calorimeter. The ``jets'' are then sorted
by energy in decreasing order. The $16$ highest energy ``jets'' are used
to provide flexible and optimized triggers based on discrimination of
individual jet energies, counting jets with energies above certain
thresholds, and determination of topological correlations between the
jets.

\section{Jet Trigger Realization}

\begin{figure*}[htb]
\centering
\includegraphics[height=12.5cm]{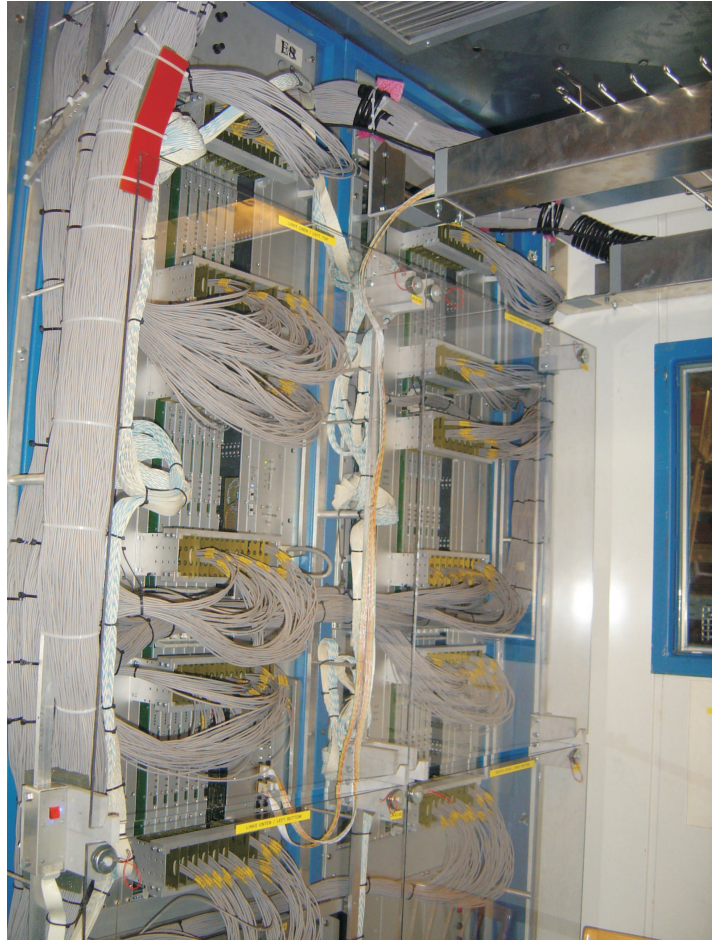}
\caption{{Details of the installed hardware in the
electronic trailer of the H1 experiment: View of the Jet Trigger ADC-Calculation-Storage unit
composed of $8$ crates, one for each of the $8$ octants of the LAr calorimeter.
The system receives the analog trigger towers and transfers the digitized signals to
the Bump Finder Unit via a bit-serial link.}}\label{fig:h1_acs}
\end{figure*}

\begin{figure*}[htb]
\centering
\includegraphics[width=11.0cm]{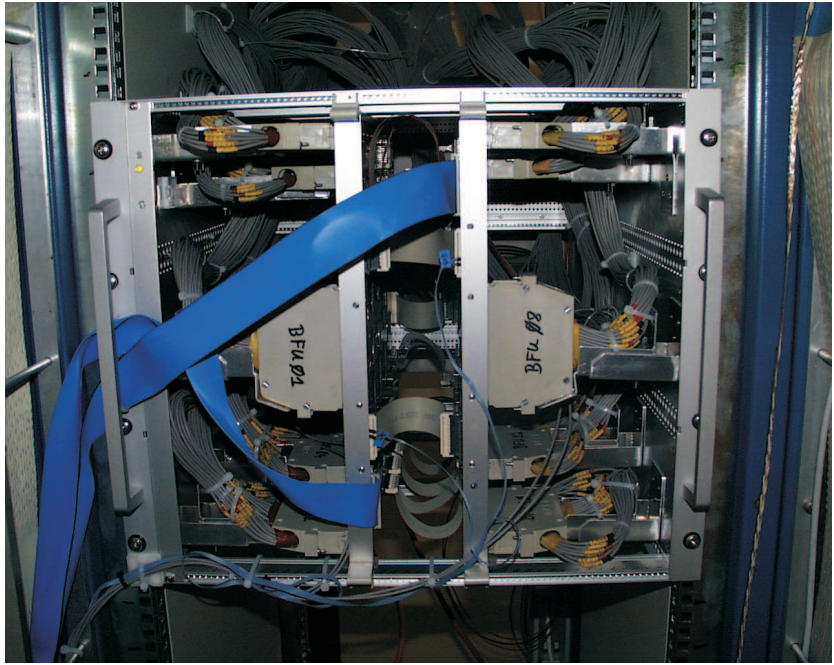}
\caption{{View into the Bump Finder crate containing 2 units, 
one for each calorimeter hemisphere. In the Bump Finder unit the ``jets'' are found in real-time. 
The Primary Sorting Unit presorts the jets from one quadrant according to their energies 
and sends its output to the Secondary Sorting Unit.}}\label{fig:h1_bfu}
\end{figure*}

\begin{figure*}[htb]
\centering
\includegraphics[width=10.0cm]{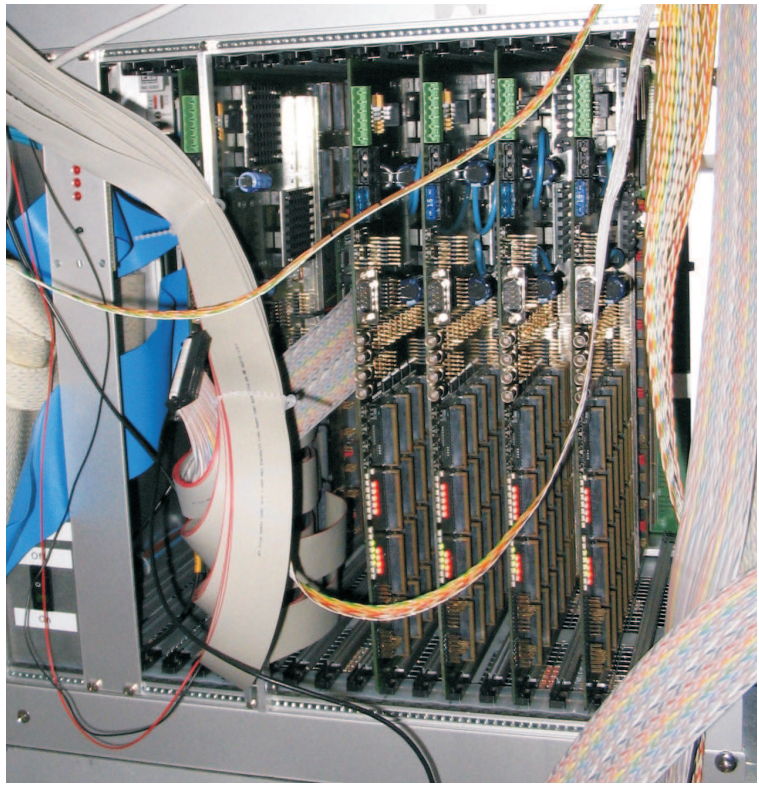}
\caption{{View into the crate housing the
Secondary Sorting Unit and the $4$ Trigger Element Generator units.  
In the Secondary Sorting Unit the presorted lists from the four quadrants are 
finally sorted by energy and transferred to the Trigger Element Generator units 
which apply topological conditions to the jets.}}\label{fig:h1_ssu}
\end{figure*}

The realization of the above algorithm was implemented in the
following way. The input of the jet trigger is $1200$ analog trigger
towers received at the $10$~MHz HERA bunch crossing rate. The clock
generation is performed by a Clock Distribution and Configuration
Card with adjustable phases to minimize the overall system latency.
The ADC-Calculation-Storage unit (see figure~\ref{fig:h1_acs}) 
digitizes the $1200$ input towers to $8$
bit accuracy each, transforms the energies into transverse energies,
and sums the electromagnetic and hadronic energies. The resulting $440$
outputs are transferred via a bit-serial link to the so-called Bump
Finder Unit (see figure~\ref{fig:h1_bfu}). 
This unit searches for local maxima of energy and sums
them with their immediate neighbors. This search and summing is done,
for each input tower, in a completely parallel fashion. The resulting
$116$ energy maxima are sorted by decreasing energy first
quadrant-wise, then detector-wise, by the Primary and Secondary
Sorting Units. The programmable Trigger Element Generator 
(see figure~\ref{fig:h1_ssu}) applies
conditions on the $16$ highest energies and their locations. These
conditions are local (energy and polar angle criteria on each
individual jet, azimuthal and polar angle differences between jets),
and global (total energy and missing energy in the event).

In total, the Jet Trigger consists of about $550$ FPGAs with $75$~M
Gates, computing $300$~G operations/s. The $12$~GB/s raw data rate is
reduced to $16$ trigger element bits per bunch cross, corresponding to
a data reduction factor of $600$. Each unit performs its function
within $1$ to $3$ bunch crossings. The total latency is $800$~ns.

\section{Jet Trigger Results}
\begin{figure*}[htb]
\centering
\includegraphics[width=13.5cm]{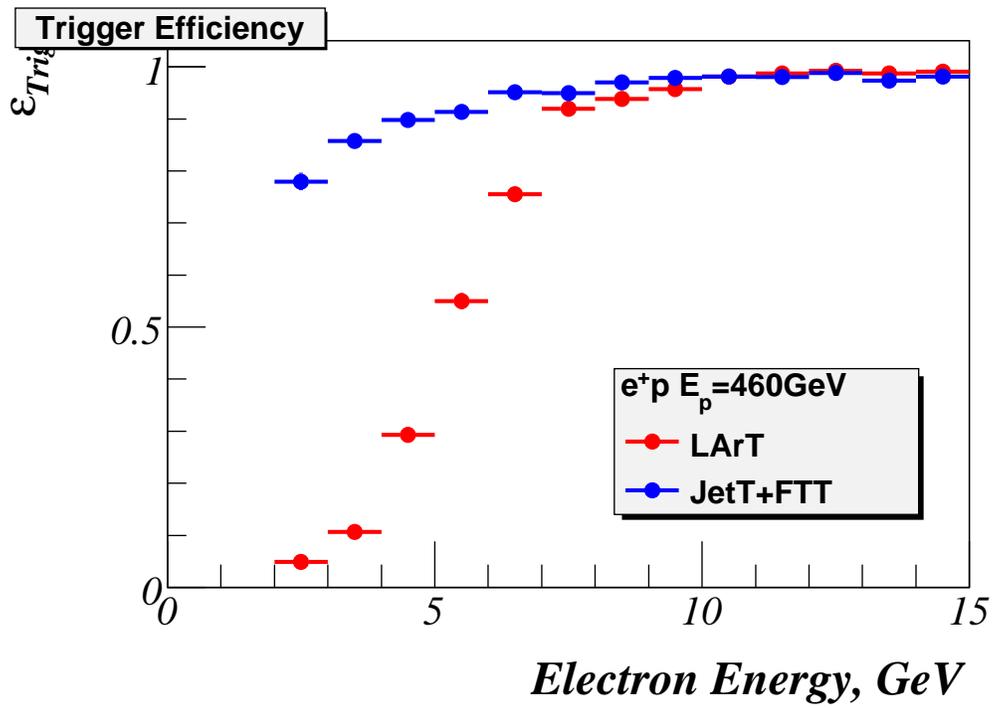}
\caption{{Efficiency to trigger on electrons as a function of the
electron energy for both the ``old'' LAr trigger (red) and the Jet Trigger (blue). 
Note the decrease of efficient triggering from $6$~GeV to $2$~GeV.}}\label{fig:H1_LowEnergy}
\end{figure*}

The Jet Trigger operation started in the summer of $2006$ and
accumulated about $100$~pb$^{-1}$ of luminosity until the end of the HERA$-2$
program in July $2007$. It opened the phase space for events containing
a single forward jet of at least $8$~GeV at low angle below $30$~degrees.
The energy-sorted jet information was combined with track-based
triggers to successfully perform b-tagging with a track threshold of
$1.5$~GeV. The Jet Trigger was used to successfully decrease the
electron triggering threshold from $6$~GeV down to $2$~GeV  
(see figure~\ref{fig:H1_LowEnergy}) and to perform
the world's first measurement of the longitudinal structure function
$F_L$ of the proton.


\end{document}